\documentclass[conference]{IEEEtran}
\IEEEoverridecommandlockouts
\usepackage{cite}
\usepackage{amsmath,amssymb,amsfonts}
\usepackage{algorithmic}
\usepackage{graphicx}
\usepackage{textcomp}
\usepackage{float}
\usepackage{xcolor}
\usepackage{tikz}
\usepackage{textcomp}
\usepackage{}
\usepackage{placeins}   % For \FloatBarrier to control figure placement
\usepackage{subcaption} % For subfigures support
\def\BibTeX{{\rm B\kern-.05em{\sc i\kern-.025em b}\kern-.08em
    T\kern-.1667em\lower.7ex\hbox{E}\kern-.125emX}}

\begin{document}

\title{Guarding Your Conversations: Privacy Gatekeepers for Secure Interactions with Cloud-Based AI Models}

\author{
    \IEEEauthorblockN{GodsGift Uzor, Hasan Al-Qudah, Ynes Ineza, Abdul Serwadda} \\
    \IEEEauthorblockA{Department of Computer Science, Texas Tech University, Lubbock, Texas\\
    Email: godsgift.uzor@ttu.edu, halqudah@ttu.edu, yineza@ttu.edu, abdul.serwadda@ttu.edu}
}

\maketitle

\begin{abstract}
The interactive nature of Large Language Models (LLMs), which closely track user data and context, has prompted users to share personal and private information in unprecedented ways. Even when users opt out of allowing their data to be used for training, these privacy settings offer limited protection when LLM providers operate in jurisdictions with weak privacy laws, invasive government surveillance, or poor data security practices. In such cases, the risk of sensitive information, including Personally Identifiable Information (PII), being mishandled or exposed remains high. To address this, we propose the concept of an "LLM gatekeeper", a lightweight, locally run model that filters out sensitive information from user queries before they are sent to the potentially untrustworthy, though highly capable, cloud-based LLM. Through experiments with human subjects, we demonstrate that this dual-model approach introduces minimal overhead while significantly enhancing user privacy, without compromising the quality of LLM responses.

%The interactive nature of Large Language Models (LLMs), which closely track user data and context, has prompted users to share personal and private information in unprecedented ways. Even when users opt out of allowing their data to be used for training, these privacy settings offer limited protection when LLM providers operate in jurisdictions with weak privacy laws, invasive government surveillance, or poor data security practices. In such cases, the risk of sensitive information—including Personally Identifiable Information (PII) --- being mishandled or exposed remains high. To address this, we propose the concept of an "LLM gatekeeper" --- a lightweight, locally run model that filters out sensitive information from user queries before they are sent to the potentially untrustworthy, though highly capable, cloud-based LLM. Through experiments with human subjects, we demonstrate that this dual-model approach introduces minimal overhead while significantly enhancing user privacy, without compromising the quality of LLM responses.
\end{abstract}

\section{Introduction}

Large Language Models (LLMs) like ChatGPT have revolutionized digital interactions by providing personalized, context-aware responses that evolve with the dialogue. Unlike traditional information sources, LLMs’ dynamic engagement often leads users to share increasingly personal details over multiple sessions, sometimes unknowingly. This gradual accumulation of sensitive information, compounded by the public's limited understanding of risks like neural network memorization, increases the likelihood of unintentional disclosure.

The issue is further exacerbated when proprietary LLMs operate in jurisdictions with weak privacy regulations, limited data security, or invasive governmental surveillance. Even opting out of data sharing for model training offers limited protection, as LLMs may still retain and link Personally Identifiable Information (PII) with sensitive details over time, posing significant privacy risks to users.
%%%%%%%%%%%%%%%%%%%%%%%%%%%%%%%%%%%%%%%%%%%%%%%%%%%%%%%%%%%%%%%%%%%%%%%%%%%%%%%%%
% To address these growing concerns, we introduce the concept of an {\it{LLM gatekeeper}}, a privacy-enhancing intermediary designed to mitigate the risk of sensitive information exposure. This gatekeeper, implemented as a lightweight, locally run model, filters out PII and other pseudo identifiers that could potentially reveal the user’s identity. It does this in real-time before the queries are transmitted to potentially untrustworthy, though highly sophisticated, cloud-based LLMs. This approach allows users to benefit from the rich knowledge and advanced capabilities of modern LLMs while maintaining stricter control over their personal information. The gatekeeper model is designed to be adaptable, allowing users to specify which types of information should be filtered, such as PII or details related to intellectual property, finance, or other sensitive domains.

% Once the queries are sanitized, they are sent via an API to the cloud-based LLM (e.g., ChatGPT), enabling users to leverage the latest technological advancements without compromising their privacy. Figure \ref{model-overview} provides a high-level representation of how this dual-model architecture operates. 
To address these privacy concerns, we introduce the {\it{LLM gatekeeper}}, a lightweight, locally run model that filters sensitive information in real time before queries reach cloud-based LLMs. Acting as a privacy-preserving intermediary, this gatekeeper detects and filters PII and other identifiers, allowing users to benefit from advanced LLMs while retaining control over personal data. It is customizable, enabling users to filter specific types of sensitive information, such as PII, intellectual property, or financial details.

Once sanitized, the queries are sent to the cloud-based LLM (e.g., ChatGPT) via an API, ensuring users can leverage powerful LLMs without sacrificing privacy. Figure \ref{model-overview} illustrates this dual-model setup, with a local model filtering user queries before they are sent to the cloud-based LLM.

%%%%%%%%%%%%%%%%%%%%%%%%%%%%%%%%%%%%%%%%%%%%%%%%%%%%%%%%%%%%%%%%%%%%%%%%%%%%%%%%%
% The idea of filtering PII to ensure privacy before submitting user data to a cloud-based system is not new. However, most previous work in this area has focused on redacting PII, which works well for data intended solely for machine learning consumption but it is problematic in real-time interactions with LLMs. In a chat session with an LLM, redactions or placeholders would disrupt the natural conversational flow and make it harder for users to interpret the LLM's responses. Our approach, therefore, not only filters out PII but also rewords the query to preserve its meaning without relying on placeholders. Additionally, our evaluations emphasize usability, an area that has not been relevant to the framing of previous works that focused on data redaction for machine learning training backends.
% In introducing and evaluating this mechanism, our paper makes the following contributions:
%While filtering PII before submitting user data to cloud-based systems is not new, most prior work focuses on redacting PII, which is effective for machine learning but problematic in real-time LLM interactions. Redactions or placeholders disrupt the natural flow of conversation, complicating user interpretation of LLM responses. Our approach addresses this by not only filtering PII but also rewording queries to retain meaning without placeholders. Additionally, we emphasize usability in our evaluations, a focus lacking in previous studies centered on data redaction for machine learning.
This paper makes the following contributions:
%%%%%%%%%%%%%%%%%%%%%%%%%%%%%%%%%%%%%%%%%%%%%%%%%%%%%%%%%%%%%%%%%%%%%%%%%%%%%%%%%
\begin{figure*}[h]
  \centering
  \includegraphics[height=0.30\linewidth]{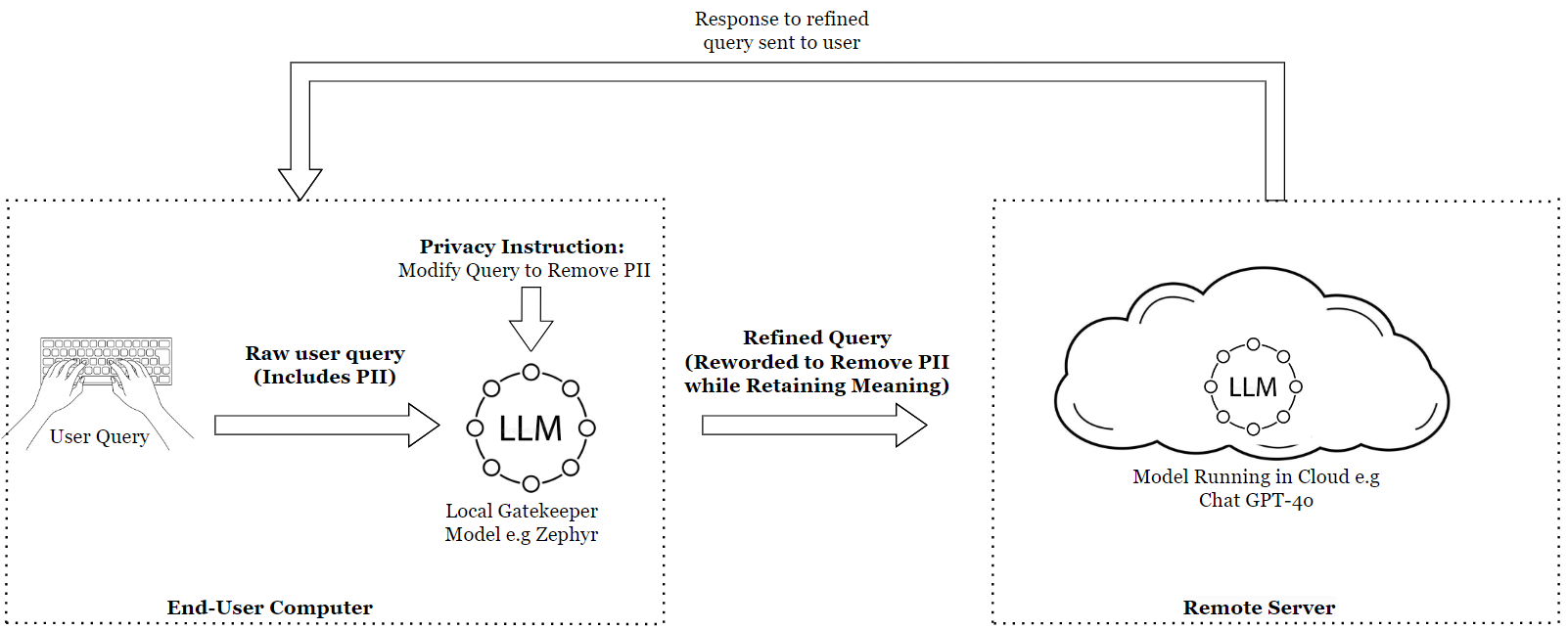}
  \caption{High-level representation of proposed gatekeeper mechanism.}
  \label{model-overview}
\end{figure*}

(1) 
% {\it{Introducing LLM gatekeepers to enhance user privacy during intersections with potentially untrustworthy cloud-based LLMs}}: We introduce the notion of a local LLM gatekeeper dedicated to filtering out private data in real-time before queries are passed on to a potentially untrustworthy, cloud-based LLM. Using two well-regarded open sourced models as the LLM gatekeeper, and ChatGPT as the proprietary cloud-based LLM, we showcase a prototype of this design. While our focus is on user privacy, it is noteworthy that this mechanism could be applied to a number of other contexts. One example is the protection of intellectual property in cases where employees rely heavily on LLMs during daily company work, yet the company cannot afford to train their own in-house LLM.
{\it{Introducing LLM gatekeepers to enhance user privacy during intersections with potentially untrustworthy cloud-based LLMs}}: We introduce a local LLM gatekeeper to filter private data in real time before forwarding queries to a cloud-based LLM, using two open-source models as the gatekeeper and ChatGPT as the cloud-based LLM in our prototype. While primarily focused on user privacy, this mechanism has broader applications, such as safeguarding intellectual property in companies where employees rely on external LLMs but lack resources to develop in-house models.
%%%%%%%%%%%%%%%%%%%%%%%%%%%%%%%%%%%%%%%%%%%%%%%%%%%%%%%%%%%%%%%%%%%%%%%%%%%%%%%%%
% (2) {\it{Experiments to characterize several key dynamics of our approach}}: We study the dynamics of our mechanism through large-scale simulations using real-world health-related questions sourced from two prominent health forums. These forums, where users regularly post sensitive personal queries, provide an authentic representation of the types of questions users are likely to ask when interacting with LLMs like ChatGPT. The diversity and length of these real queries also offer a realistic assessment of our system’s performance in scenarios resembling actual user interactions.

% By processing these real-world questions through our dual-model setup, we systematically evaluate several key metrics, including the additional processing time introduced by the gatekeeper model compared to a traditional single-cloud LLM setup, the semantic similarity between the original and refined queries, and the similarity between the final response and that of a single cloud-based model. Our findings provide critical insights into how well our system preserves the intent of user queries and serve as a foundational benchmark for our next contribution, where we incorporate end-user involvement into the experiment.

(2) {\it{Experiments to characterize several key dynamics of our approach}}:
 We analyze our mechanism with large-scale simulations using real health-related questions sourced from prominent forums. These questions, often sensitive and detailed, closely resemble the queries users might pose to LLMs like ChatGPT, providing a realistic assessment of system performance in authentic scenarios.

Through our dual-model setup, we evaluate key metrics, such as processing time added by the gatekeeper, semantic alignment between original and refined queries, and similarity in responses compared to a single cloud-based model. These findings offer some insight into how effectively our system preserves query intent and informs our next phase of user-centered evaluation.

(3) {\it{User studies on the need and efficacy of this scheme}}: We conducted a user study aimed at evaluating the practical usability of our solution, prompting its design given the perceived risks. 
%The first study focuses on users' concerns regarding the potential leakage of private information through cloud-based proprietary LLMs. Participants completed a survey with questions covering a range of privacy issues associated with LLM usage. 
The study tested the functionality of our prototype, in which users interacted with it using their own queries. They provided feedback on several aspects, including response times, accuracy in preserving the intended meaning of queries, and the effectiveness of the gatekeeper in filtering out sensitive information deemed private by the users. This study provided comprehensive information on the need and efficacy of our proposed privacy-enhancing solution.

\section{Threat Model}
\label{threat}
%%%%%%%%%%%%%%%%%%%%%%%%%%%%%%%%%%%%%%%%%%%%%%%%%%%%%%%%%%%%%%%%%%%%%%%%%%%
% This section prepares the ground for our design and experiments by laying out the threat landscape that necessitates our solution. We describe the type of user most vulnerable to these threats, the specific risks they face, and the adversaries and technical vulnerabilities that make cloud-based LLMs a potential privacy risk.

This section outlines the threat landscape that motivates our design and experiments. We identify the types of users most vulnerable to privacy risks, detail the specific threats they face, and describe the adversaries and technical vulnerabilities that make cloud-based LLMs a potential privacy risk.
%%%%%%%%%%%%%%%%%%%%%%%%%%%%%%%%%%%%%%%%%%%%%%%%%%%%%%%%%%%%%%%%%%%%%%%%%%%%%%
\subsection{Motivation and Context}
% Our solution is applicable to all users of cloud-based LLM models, but it is particularly beneficial for the growing group of users who have become heavily reliant on LLMs for everyday personal and professional tasks. The deep understanding LLMs provide, along with their ability to remember and track conversations over multiple sessions, has made them essential tools in many users' workflows. These users often turn to LLMs for sensitive matters, such as health-related inquiries, issues concerning proprietary company data, and personal documents like scanned IDs or financial records.

% During these interactions, users may, knowingly or unknowingly, share Personally Identifiable Information (PII) or pseudo-identifiers. While these pseudo-identifiers may not reveal the user’s identity on their own, they can be combined with other data or publicly available information to do so. Over time, this data builds up, creating a detailed profile of the user. If this information, especially when combined with PII, were to be leaked or misused, the consequences could be severe, ranging from reputational damage to strained personal relationships, legal trouble, or financial loss.

Our solution serves all cloud-based LLM users, but is particularly advantageous for those heavily reliant on LLMs for personal or professional tasks. These users frequently turn to LLMs for sensitive matters, such as health inquiries, proprietary company data, or personal documents (e.g., IDs, financial records). During these interactions, they may disclose PII or pseudo-identifiers, which, while not revealing identity alone, can be combined with other data to do so. This accumulation over time creates a detailed user profile, where potential leaks or misuse could lead to severe consequences, including reputational harm, strained relationships, legal issues, or financial loss.
%%%%%%%%%%%%%%%%%%%%%%%%%%%%%%%%%%%%%%%%%%%%%%%%%%%%%%%%%%%%%%%%%%%%%%%%%%%%%%%%%
\subsection{Adversaries and Technical Risks}
In this model, the user must guard against a range of adversaries and technical risks, including:

% {\bf{Cloud LLM Operators}}: 
% The cloud-based AI platforms relevant to our work can be categorized into two main groups: those that cannot be trusted due to the political or regulatory environments they operate in, and those that may be otherwise trustworthy but still vulnerable to technical failures like model memorization. For providers operating in jurisdictions with weak privacy laws, invasive surveillance, or repressive governments, the risk is that sensitive user data might be harvested, monitored, or shared with third parties, including advertisers or government agencies. These operators may act unethically or be compelled to comply with oppressive regulations, putting users at significant risk.

% On the other hand, even trustworthy providers with good intentions face the inherent risk of model memorization, where the model inadvertently retains and recalls sensitive information. Memorized data could be leaked to other users through responses, despite the provider following ethical guidelines and prioritizing user privacy. 

{\bf{Cloud LLM Operators}}: Cloud-based AI platforms fall into two main risk categories: those untrustworthy due to political or regulatory constraints, and those vulnerable to technical issues, such as model memorization. Platforms in regions with weak privacy laws, invasive surveillance, or repressive governments may expose sensitive user data to unauthorized monitoring or third-party access, including advertisers and government agencies. Such operators may act unethically or be compelled by regulation, significantly endangering user privacy.

Even trustworthy providers face the risk of model memorization, where sensitive information is unintentionally retained and recalled in responses, potentially exposing user data to others. This risk persists despite ethical practices and a commitment to user privacy.
%%%%%%%%%%%%%%%%%%%%%%%%%%%%%%%%%%%%%%%%%%%%%%%%%%%%%%%%%%%%%%%%%%%%%%%%%%%%%%%%%

% {\bf{Insider Threats}}: Employees or contractors working for the cloud platform might access sensitive data for personal gain or malicious intent. The scale of data processing by these platforms makes it difficult to guarantee that all individuals with access to the data will handle it responsibly.
{\bf{Insider Threats}}: Employees or contractors of cloud platforms may access sensitive data for personal or malicious purposes. Given the scale of data processing, ensuring responsible handling by all individuals with access is challenging.
%%%%%%%%%%%%%%%%%%%%%%%%%%%%%%%%%%%%%%%%%%%%%%%%%%%%%%%%%%%%%%%%%%%%%%%%%%%%%%%%%
{\bf{External Hackers}}: Cloud-based services are often targets for cyberattacks, which can lead to the theft of large volumes of sensitive user data. Whether through vulnerabilities in the system or advanced attacks, there is always a risk that sensitive information might be exposed through external breaches.

{\bf{Third-Party Data Requests or Subpoenas}}: In certain jurisdictions, government agencies or other third parties can compel cloud operators to hand over user data through legal orders, without the user’s knowledge. This is particularly concerning in countries with poor human rights records or invasive surveillance practices.

\subsection{Local Model and Limitations}
%%%%%%%%%%%%%%%%%%%%%%%%%%%%%%%%%%%%%%%%%%%%%%%%%%%%%%%%%%%%%%%%%%%%%%%%%%%
% The user has access to an open-source model that can be run locally on their computer. Ideally, the user would rely solely on this model for all LLM needs, avoiding the risks of using an untrustworthy cloud-based LLM. However, several limitations prevent the local model from serving as a comprehensive solution, restricting its role to that of a gatekeeper. These limitations include:

Users could ideally rely entirely on an open-source, locally-run model to meet all LLM needs, thereby avoiding the risks of untrusted cloud-based LLMs, several limitations restrict this model to a gatekeeper role. These limitations include:
%%%%%%%%%%%%%%%%%%%%%%%%%%%%%%%%%%%%%%%%%%%%%%%%%%%%%%%%%%%%%%%%%%%%%%%%%%%%%
{\bf{Reduced Capability}}: Open-source models running locally are generally smaller and less powerful than their cloud-based counterparts, which benefit from continuous updates and optimization through vast computational resources. Consequently, local models might struggle with delivering high-quality and accurate responses, particularly for complex or specialized queries.

{\bf{Hardware Constraints}}: Running large language models locally requires substantial computing power, typically high-performance GPUs and significant memory. Many users may lack the hardware necessary to support more advanced models, forcing them to use smaller, less capable versions. While these models may handle simpler tasks like query rewriting, they can fall short when tackling more demanding tasks, reducing their overall effectiveness.

{\bf{Limited Access to Latest Data}}: Unlike cloud-based models, which are regularly updated with the latest information across a wide range of fields, local models can become outdated unless the user actively maintains them. This lack of regular updates can impair the model's ability to provide accurate or relevant information, particularly for fast-evolving domains such as healthcare or finance.
\section{Related Research}
\label{related1}
\subsection{LLM-Driven Data Sanitization}
%%%%%%%%%%%%%%%%%%%%%%%%%%%%%%%%%%%%%%%%%%%%%%%%%%%%%%%%%%%%%%%%%%%%%%%%%
% The sanitization of user data prior to submission to machine learning systems and related technologies in the cloud is well-established as one of the layers of defense against data privacy breaches. However, the rise of LLMs has brought renewed attention to this issue for two critical reasons. First, LLMs excel at detecting PII and other sensitive data that traditional rule-based methods may miss, driving interest in leveraging LLM-based solutions to strengthen privacy safeguards. Second, from an end-user perspective, LLMs significantly increase privacy risks by augmenting user input data with patterns learned from vast datasets, including public web pages and social media content that were included in their training sets, to make potentially harmful inferences about users. As a result, there is a growing body of research focused on developing LLM-oriented data sanitation tools, motivated by the inherent privacy risks posed by LLMs themselves. Below, we reference a selection of recent representative works in this area and outline how our approach diverges from these existing solutions.

Sanitizing user data before submission to machine learning systems has long protected against privacy breaches, but LLMs bring new concerns. First, LLMs excel at detecting PII beyond traditional rule-based methods, enhancing privacy potential. However, LLMs also pose risks by embedding patterns from extensive datasets—such as public web content—which can lead to unintended inferences about users. Consequently, there is increasing research into LLM-specific data sanitization tools to address these unique privacy challenges. Below, we review recent works and highlight how our approach diverges from existing solutions.
%%%%%%%%%%%%%%%%%%%%%%%%%%%%%%%%%%%%%%%%%%%%%%%%%%%%%%%%%%%%%%%%%%%%%%%%%%%%%%%%
% Kim et al. built ProPILE  \cite{Kim1}, a tool that allows users to assess the potential leakage of their PII in LLMs. With only black-box access, users can send prompts to test for PII leakage. ProPILE is also valuable for LLM service providers to evaluate and mitigate privacy vulnerabilities. Experiments on Open Pre-trained Transformers (OPT) confirm that significant PII can be leaked through refined prompts. ProPILE aims to raise awareness about PII risks and help improve LLM privacy measures. 

Kim et al. developed ProPILE \cite{Kim1}, a tool for assessing PII leakage in LLMs using black-box access. Users can test prompts for PII exposure, and LLM providers can leverage ProPILE to identify and mitigate privacy vulnerabilities. Experiments with Open Pre-trained Transformers (OPT) demonstrate that refined prompts can reveal significant PII, highlighting ProPILE’s role in raising awareness and advancing LLM privacy.
%%%%%%%%%%%%%%%%%%%%%%%%%%%%%%%%%%%%%%%%%%%%%%%%%%%%%%%%%%%%%%%%%%%%%%%%%
% The work in \cite{10433801} used GPT-4 to mask PII in health-oriented data that was synthetically generated by the GPT-3.5 model. The data was generated in such a way to represent 1000 patients, 20 doctors and 100 different illnesses. Their PII masking design employed a mapping table that consistently replaced each unique instance of PII with a pre-determined string (e.g., the word, malaria was always replaced by a certain specific string). They measured the success of their scheme by counting the numbers of patients per doctor, patients diagnosed with specific illnesses, and distribution of illnesses across doctors. The tool registered very high success rates on all these metrics.

In \cite{10433801}, GPT-4 was used to mask PII in synthetic, health-oriented data generated by GPT-3.5 to represent 1,000 patients, 20 doctors, and 100 illnesses. Their method employed a mapping table to consistently replace each unique PII instance with a predefined string (e.g., "malaria" was always replaced by the same string). Success was measured by tracking patients per doctor, illness counts, and illness distribution across doctors, achieving high accuracy on all metrics.
%%%%%%%%%%%%%%%%%%%%%%%%%%%%%%%%%%%%%%%%%%%%%%%%%%%%%%%%%%%%%%%%%%%%%%%%%%%%%%%
% In \cite{chong2024}, Chong et al.  introduces a tool designed to protect user privacy in real-time when interacting with cloud-based LLM services. It uses a three-layered sanitization process to redact PII before prompts are sent to LLM providers. The first layer is a rule-based filter which redacts sensitive information based on pre-defined keywords and generic matching rules. The second is a machine learning (ML)-based detector that identifies named entities, such as names, locations, and organizations. Finally, they use a local LLM that identifies privacy-sensitive topics in the prompts and warns users about potential privacy risks. They evaluated the tool based on queries synthetically generated by an LLM, and showed that it achieved high accuracy in detecting PII (98.5\%) and privacy-sensitive topics (89.9\%) while maintaining a low performance overhead.

In \cite{chong2024}, Chong et al. introduce a privacy tool for real-time PII sanitization in cloud-based LLM interactions. The tool uses a three-layered approach: a rule-based filter for redacting sensitive information via keywords, an ML-based detector for named entities (e.g., names, locations), and a local LLM for identifying privacy-sensitive topics and alerting users. Tested on synthetically generated queries, the tool demonstrated high accuracy in PII detection (98.5\%) and topic identification (89.9\%) with minimal performance overhead.
%%%%%%%%%%%%%%%%%%%%%%%%%%%%%%%%%%%%%%%%%%%%%%%%%%%%%%%%%%%%%%%%%%%%%%%%%%%%%%%%%
% The work in \cite{Anthi}, presents a text sanitization approach that addresses the limitations of traditional methods that rely on predefined entity categories. It introduces a privacy-enhanced entity recognizer capable of detecting personal information beyond named entities, such as demographic attributes. By integrating empirical privacy risk measures, it optimizes which entities to mask, balancing privacy and data utility.

The work in \cite{Anthi} presents a text sanitization method that extends beyond traditional entity-based approaches by detecting a wider range of personal information, including demographic attributes. By integrating empirical privacy risk measures, it selectively masks entities to optimize the balance between privacy and data utility.
%%%%%%%%%%%%%%%%%%%%%%%%%%%%%%%%%%%%%%%%%%%%%%%%%%%%%%%%%%%%%%%%%%%%%%%%%%%%%%%

% The work by Wiest et al. \cite{Wiest2024} presents an anonymization tool, the LLM-Anonymizer, which uses large language models (LLMs) to de-identify patient information in medical documents while maintaining privacy, addressing challenges in sharing real-world medical data for research. The researchers benchmarked eight locally deployable LLMs, including Llama-3 and Llama-2 models, on 100 clinical letters to extract and remove personal identifying information (PII). The LLM-Anonymizer achieved high accuracy, with the Llama-3 70B model reaching 98.05\% accuracy in de-identifying sensitive data, missing only 1.95\% of PII.

Wiest et al. \cite{Wiest2024} developed the LLM-Anonymizer, a tool leveraging large language models (LLMs) to de-identify patient information in clinical documents, facilitating data sharing for research while preserving privacy. Testing eight locally deployable LLMs, including Llama-3 and Llama-2 models, on 100 clinical letters, the LLM-Anonymizer showed strong results, with the Llama-3 70B model achieving 98.05\% accuracy in PII detection, leaving only 1.95\% of sensitive information unmasked.
%%%%%%%%%%%%%%%%%%%%%%%%%%%%%%%%%%%%%%%%%%%%%%%%%%%%%%%%%%%%%%%%%%%%%%%%%%%%%%

\subsection{How We Differ From Prior Research}
Our research differs from the above research in several significant ways:
%%%%%%%%%%%%%%%%%%%%%%%%%%%%%%%%%%%%%%%%%%%%%%%%%%%%%%%%%%%%%%%%%%%%%%%%%%%%%%%
% {\bf{Human Subjects Experiments}}: A key distinguishing factor in our approach is the inclusion of human subject experiments, where users interact with the system and provide feedback on various performance attributes, such as usability, response quality, and effectiveness in protecting privacy. None of the previously discussed works incorporate this kind of user-centered evaluation, which we believe is crucial for assessing how well the system performs in real-world scenarios.

{\bf{Human Subjects Experiments}}: Our approach uniquely includes human subject experiments, allowing users to interact with the system and provide feedback on usability, response quality, and privacy protection. This user-centered evaluation, absent in prior studies, is crucial for assessing real-world system performance.
%%%%%%%%%%%%%%%%%%%%%%%%%%%%%%%%%%%%%%%%%%%%%%%%%%%%%%%%%%%%%%%%%%%%%%%%%%%%%%%%%

% {\bf{Rewording Instead of Redacting}}: All the aforementioned systems primarily focus on redacting user queries by using placeholders to mask sensitive information. While this is effective for data meant solely for machine learning models, it creates challenges in real-time user interactions, where even a few placeholders can quickly confuse the conversation. Our approach filters PII and rewords the query to maintain the natural flow of the conversation, prioritizing usability. This emphasis on user-friendly interaction sets our work apart from these systems.
{\bf{Rewording Instead of Redacting}}: Prior systems mostly rely on redacting sensitive information with placeholders, which, while effective for machine learning tasks, can disrupt real-time user interactions by creating confusing gaps. Our approach, by contrast, rephrases queries to remove PII while preserving conversational flow, enhancing usability and setting our work apart.
%%%%%%%%%%%%%%%%%%%%%%%%%%%%%%%%%%%%%%%%%%%%%%%%%%%%%%%%%%%%%%%%%%%%%%%%%%%%%%%%%

% {\bf{Evaluation Metrics Focused on Semantics}}: As a result of our rewording approach, our performance evaluation differs significantly from the above works, which typically assess the effectiveness of placeholder replacements by counting how many instances of a certain PII element were successfully redacted (e.g., see \cite{10433801, chong2024}). In contrast, a key part of our evaluation focuses on preserving the contextual and language semantics of the reworded queries. To achieve this, we use text embeddings and similarity metrics to ensure that the reworded queries retain their original meaning, providing a more nuanced assessment than simple placeholder replacement.

{\bf{Evaluation Metrics Focused on Semantics}}: Our rewording approach leads to a distinct evaluation method compared to prior works that measure placeholder replacement accuracy for redacted PII (e.g., \cite{10433801, chong2024}). Instead, we assess how well reworded queries preserve original context and semantics, using text embeddings and similarity metrics. This enables a more nuanced evaluation, focused on meaning retention rather than mere placeholder substitution.
%%%%%%%%%%%%%%%%%%%%%%%%%%%%%%%%%%%%%%%%%%%%%%%%%%%%%%%%%%%%%%%%%%%%%%%%%%
% {\bf{Synthetic vs. Real Data}}: Several of the above studies rely on synthetic data to evaluate their schemes (e.g., see \cite{10433801, chong2024}), often using one LLM to generate data and another to sanitize it. This creates uncertainty about how the LLM dependencies affect the outcomes, as an LLM might exploit LLM-like patterns in the synthetic data that wouldn't exist in human-generated content. Our work, by contrast, uses only human-generated data in all evaluations. One part of our evaluation involves simulating interactions using real-world data from health platforms, where users ask genuine questions. The other part involves actual human participants who interacted with the system and submitted their own queries. This reliance on human-generated data ensures that our results are more reflective of how the system would perform in real-world use cases, free from LLM dependencies in the datasets used for testing.

{\bf{Synthetic vs. Real Data}}: Many or the prior studies (e.g., \cite{10433801, chong2024}) use synthetic data often generated by one LLM and sanitized by another, raising questions about LLMspecific dependencies that may not mirror real-world contexts. In contrast, our evaluation relies solely on human-generated data: real health platform interactions and live queries from human participants. This approach provides more accurate insights into real-world system performance, minimizing dependencies from LLM-synthesized data.

%%%%%%%%%%%%%%%%%%%%%%%%%%%%%%%%%%%%%%%%%%%%%%%%%%%%%%%%%%%%%%%%%%%%%%%%%%%%%
% {\bf{Real-time Design and Evaluation}}: Except for the work in \cite{chong2024}, none of the above mechanisms were designed for the use case of a real-time system acting as an intermediary between the user and the cloud-based LLM. In fact, even the system in \cite{chong2024} was evaluated using simulated data read from a CSV file, which lacks the dynamics of live user interaction. In contrast, our work is explicitly designed and evaluated using real-time user interactions. This focus on actual real-time performance distinguishes our evaluations and findings from past works, which were centered on static data processing.

{\bf{Real-time Design and Evaluation}}: Except for \cite{chong2024}, none of the prior mechanisms were designed for real-time intermediary use between a user and a cloud-based LLM. Even in \cite{chong2024}, evaluation was limited to simulated CSV data, lacking live interaction dynamics. In contrast, our work is explicitly built for and tested in real-time user interactions, distinguishing our approach from prior static data evaluations.
%%%%%%%%%%%%%%%%%%%%%%%%%%%%%%%%%%%%%%%%%%%%%%%%%%%%%%%%%%%%%%%%%%%%%%%%%%%%%%%%
% \begin{table}[]
% \centering
% \resizebox{\columnwidth}{!}{%
% \begin{tabular}{|l|l|l|l|l|}
% \hline
% \textbf{Models} & \textbf{Release Date} & \textbf{Size (GB)} & \textbf{Organization} & \textbf{\begin{tabular}[c]{@{}l@{}}Used as Gatekeeper\\ (Yes/No)\end{tabular}} \\ \hline
% Phi3.5 \cite{phi3.5} & April 2024 & 2.2 & Microsoft & Yes \\ \hline
% Llama2 \cite{touvron2023llama} & July 2023 & 3.8 & Meta & No \\ \hline
% Code Llama \cite{Rozire2023CodeLO} & August 2023 & 3.8 & Meta & No \\ \hline
% Starling \cite{starling2023} & November 2023 & 4.1 & Banghua et al. & No \\ \hline
% Mistral \cite{Jiang2023Mistral7} & September 2023 & 4.1 & Mistral AI & No \\ \hline
% Zephyr \cite{tunstall2023zephyr} & November 2023 & 4.1 & \begin{tabular}[c]{@{}l@{}}Web Pilot\\ Hugging Face\end{tabular} & No \\ \hline
% Llama3.1 \cite{dubey2024llama} & July 2024 & 4.7 & Meta & No \\ \hline
% Gemma2 \cite{gemma_2024} & July 2024 & 5.4 & Google & Yes \\ \hline
% \end{tabular}%
% }
% \caption{Popular open-source LLMs from which the gatekeepers used in our experiments were chosen.}
% \label{table:example}
% \end{table}

\begin{table}[]
\centering
\resizebox{\columnwidth}{!}{%
\begin{tabular}{l l l l l}
\hline
\textbf{Models} & \textbf{Release Date} & \textbf{Size (GB)} & \textbf{Organization} & \textbf{\begin{tabular}[c]{@{}l@{}}Used as Gatekeeper\\ (Yes/No)\end{tabular}} \\ \hline
Phi3.5 \cite{phi3.5} & April 2024 & 2.2 & Microsoft & Yes \\ \hline
Llama2 \cite{touvron2023llama} & July 2023 & 3.8 & Meta & No \\ \hline
Code Llama \cite{Rozire2023CodeLO} & August 2023 & 3.8 & Meta & No \\ \hline
Starling \cite{starling2023} & November 2023 & 4.1 & Banghua et al. & No \\ \hline
Mistral \cite{Jiang2023Mistral7} & September 2023 & 4.1 & Mistral AI & No \\ \hline
Zephyr \cite{tunstall2023zephyr} & November 2023 & 4.1 & \begin{tabular}[c]{@{}l@{}}Web Pilot\\ Hugging Face\end{tabular} & No \\ \hline
Llama3.1 \cite{dubey2024llama} & July 2024 & 4.7 & Meta & No \\ \hline
Gemma2 \cite{gemma_2024} & July 2024 & 5.4 & Google & Yes \\ \hline
\end{tabular}%
}
\caption{Popular open-source LLMs from which the gatekeepers used in our experiments were chosen.}
\label{table:example1}
\end{table}

% \begin{figure*}[ht]
%   \centering
%   % \includegraphics[height=0.38\linewidth]{userinterface.png}
%   \includegraphics[width=\textwidth]{userinterface_query.png}
%   %\includegraphics[width=\textwidth]{LLM2.png}
%   \caption{User-interface for our prototype. A sample user query (bottom section), refined query (top section), and final response from the combined LLM pipeline (middle section) are shown on the interface.}
%   \label{user-interface1}
% \end{figure*}
\begin{figure*}[ht]
  \centering
  \includegraphics[width=\textwidth, height=0.40\linewidth]{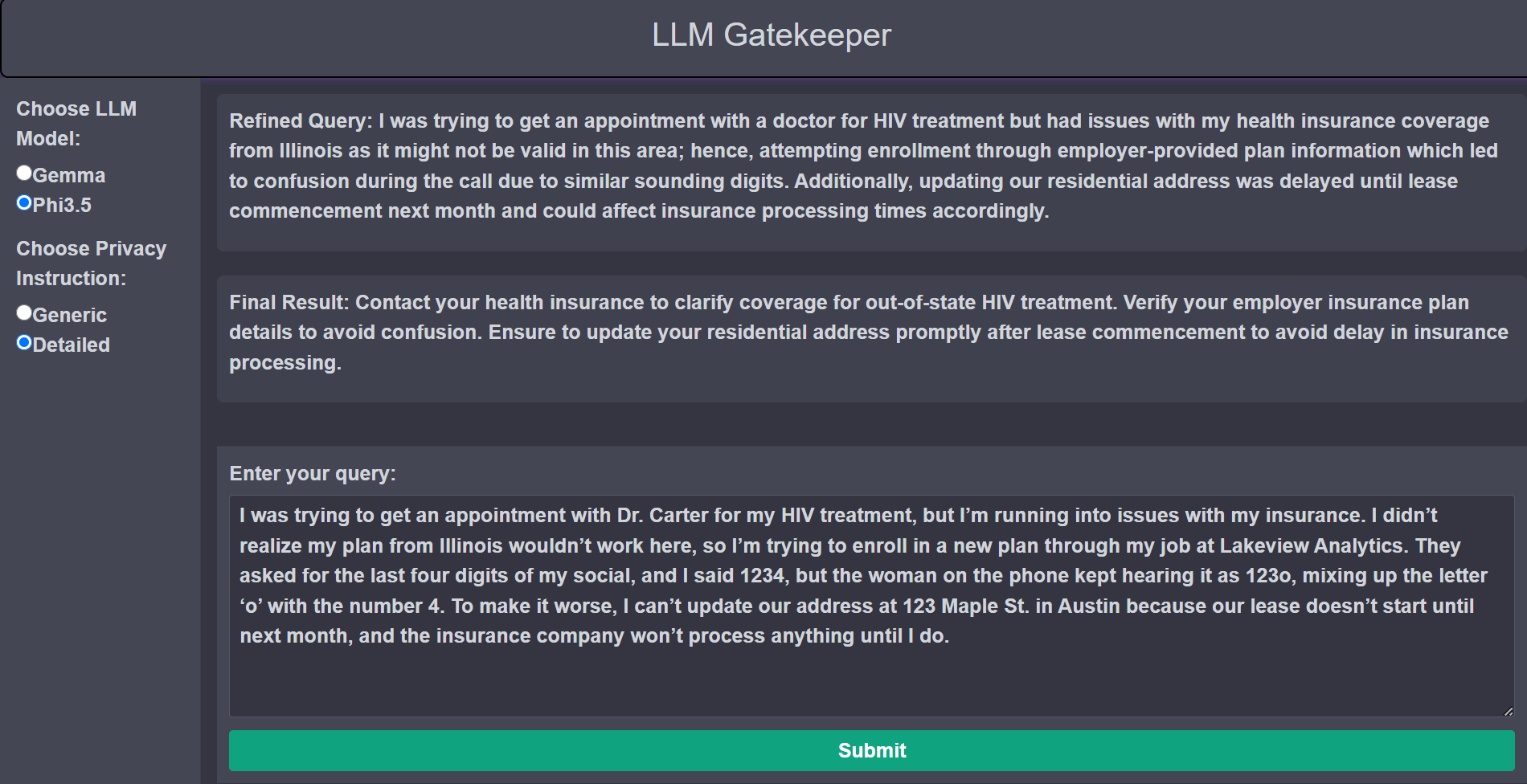}
  \caption{User-interface for our prototype. A sample user query (bottom section), refined query (top section), and final response from the combined LLM pipeline (middle section) are shown on the interface.}
  \label{user-interface1}
\end{figure*}

\section{Experimental Design}
\label{sec:Experimental-Design}

\subsection{Prototype Implementation}
\label{sec:Prototype-Implementation}
Our prototype was implemented as a Python Flask application, designed to route user queries through a gatekeeper model before forwarding them to ChatGPT (specifically GPT-4o) via the API. The gatekeeper refines the original query by filtering out Personally Identifiable Information (PII), creating a sanitized version that is then passed along to ChatGPT. This gatekeeper model runs locally using the Ollama platform \cite{ollama}, which supports running large language models (LLMs) on local systems.

The prototype offers two gatekeeper options: Gemma2 \cite{gemma_2024} and Phi3.5 \cite{phi3.5}, selected from a broader set of models (see Table \ref{table:example1}). Our selection criteria focused on balancing model recency and variability in complexity and size. We chose one lightweight model, suitable for users with limited resources, and one more complex model for users without hardware constraints. This comparison allows us to assess performance differences in terms of answer quality, PII filtering effectiveness, and response time across these two extremes. As seen in Table \ref{table:example1}, Gemma2 represents the high-end model, while Phi3.5 simulates the low-resource scenario.

The user interface (Figure \ref{user-interface1}) allows users to select a Privacy Instruction from two options: a generic or a detailed instruction. This setting addresses the phenomenon of prompt engineering \cite{aws_prompt_engineering}, where LLM performance is highly dependent on the structure of the prompt. The generic instruction offers high-level guidance for removing PII, while the detailed instruction explicitly defines the types of information to be filtered. We compare the performance of these two strategies to inform the design of future gatekeeper models. Below was our first version for both instruction texts. 

\noindent\textbf{Generic Privacy Instruction:}
\begin{quote}
\textit{Please rewrite the provided text to remove Personally Identifiable Information (PII) while keeping the meaning of the text unchanged.}
\end{quote}

\noindent\textbf{Detailed Privacy Instruction:}
\begin{quote}
\textit{Rewrite the provided text to remove Personally Identifiable Information (PII) while keeping the meaning of the text unchanged. Examples of PII include names, social security numbers, driver’s license numbers, etc., that uniquely identify individuals. PII also includes information that, while not uniquely identifying on its own, could be combined with other details to identify individuals or link them to sensitive information.}
\end{quote}

During our preliminary testing, we observed that the models sometimes produced overly verbose intermediate queries, sometimes adding unsolicited extra details, which could obscure the user’s intended meaning. To mitigate this, we modified the detailed instruction by adding:
"Do not make the response verbose. Only provide the refined query. Do not include any additional information."

The user interface for our prototype includes a text area for the user to type their query, as well as text areas to visualize the refined query after PII filtering, and the final response from ChatGPT. For all experiments, we deployed the prototype on a Dell desktop with the following specifications: 12th Gen Intel(R) Core(TM) i7-12700k 3.60GHz processor, 64GB RAM, a 24.0 GB NVIDIA GeForce RTX 3090 GPU, running on a Windows 11 Professional 64-bit operating system.

% \begin{figure*}[htbp]
%   \centnmering
%   % First Subfigure
%   \begin{subfigure}[t]{0.25\textwidth}
%     \centering
%     \includegraphics[width=\textwidth]{CDF_long.png}
%     \caption{Healthtap Dataset}
%     \label{fig:cdflong}
%   \end{subfigure}
%   \hspace{0.05\textwidth} % Adjust space between figures if needed
%   % Second Subfigure
%   \begin{subfigure}[t]{0.25\textwidth}
%     \centering
%     \includegraphics[width=\textwidth]{CDF_short.png}
%     \caption{MeQSum Dataset}
%     \label{fig:cdfshort}
%   \end{subfigure}
%   \caption{CDFs of Query Sizes in Our Simulation Datasets.}
%   \label{fig:cdf}
% \end{figure*}
\begin{figure}
    \centering
    \includegraphics[width=\columnwidth]{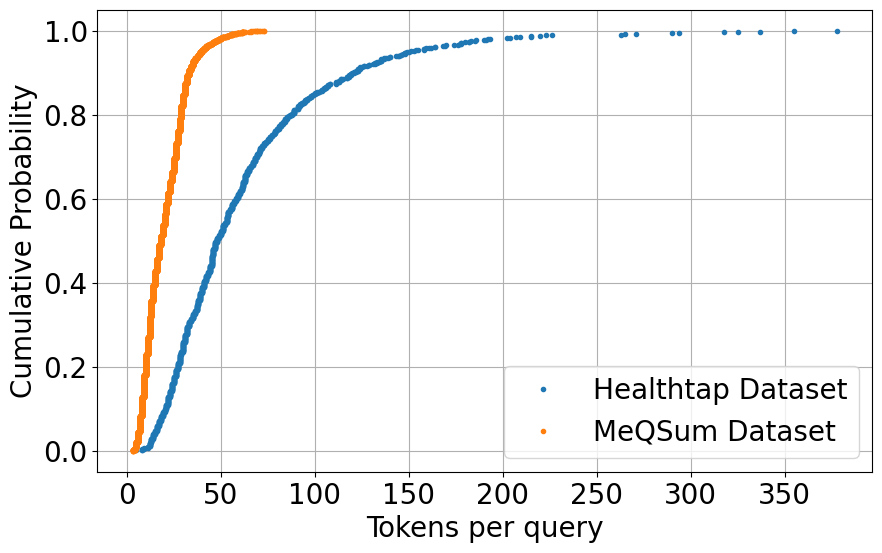}
    \caption{CDFs of Query Sizes in Our Simulation Datasets.}
    \label{fig:cdf}
\end{figure}

\subsubsection{Motivation Behind Simulations}
While the ultimate test for a tool like ours is direct user interaction, the number of tests that users can reasonably tolerate is inherently limited due to the time and cognitive load required for generating thoughtful queries and providing detailed feedback. For instance, a user is unlikely to tolerate creating a large number of queries, especially if they are required to be  large or complex. These kinds of tests, however, are crucial for thoroughly understanding the system’s behavior.

To address these limitations, we conducted a series of simulated tests to gather insights on system dynamics before initiating user experiments. This allowed us to explore how the system behaves under different conditions, such as large or methodically varying query sizes, without overburdening users. Additionally, this preliminary testing enabled us to identify and resolve key issues (such as the verbose responses mentioned in Section \ref{sec:Prototype-Implementation}) before proceeding to the user studies.

\subsubsection{Data Sources for Simulations} For these simulations to realistically represent system behavior, careful attention must be given to the nature of the queries used. While it might be tempting to generate queries using an LLM, this approach risks producing queries that do not accurately reflect real human behavior. The validity of the experiment would then be questionable, especially considering that we are using an LLM for evaluations as well. To mitigate this, we used actual user queries from two well-known health forums, ensuring the queries are authentic and representative of real-world interactions.

The first dataset we used was the Medical Question Pairs (MQP) dataset, which contains 3,048 medical questions from HealthTap, a telehealth provider platform \cite{mccreery2020effective}. The second dataset was the Medical Question Summarization (MeQSum) corpus \cite{MeQSum}, which comprises 1,000 summarized consumer health questions. Figure \ref{fig:cdf} shows the distribution of query sizes (counted in tokens) found in the two datasets. These CDFs provide a guide for choice of representative query sizes during the simulations. For each of 4 query size ranges selected from across the full size spectrum seen on these CDFs, we randomly selected 30 queries from these datasets. The specific token sizes that we used are shown below:
\begin{figure}[ht]
    \centering
    \includegraphics[width=\columnwidth]{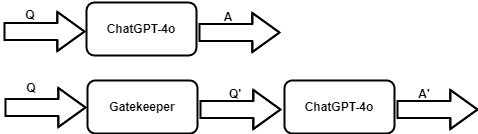}
    \caption{Comparing ChatGPT to the Gatekeeper system. $Q$ is the user query, while $A$ is ChatGPT's response. $Q'$ is the gatekeeper's refined query given the same input $Q$, while $A'$ is the final output from our full privacy-enhancing pipeline.}
    \label{fig:answer-consistency}
\end{figure}
\begin{itemize}
    \item Range 25 - 40 tokens * 30 queries
    \item Range 50 - 80 tokens * 30 queries
    \item Range 100 - 160 tokens * 30 queries
    \item Range 200 - 320 tokens * 30 queries
\end{itemize}

Observe that we selected the query sizes in an exponentially increasing pattern, doubling the range each time. This approach was chosen because small linear shifts in the query size would likely not result in significant changes in LLM behavior unless a large number of such shifts were tested, which would make the experiment inefficient. We hypothesized that larger, more aggressive shifts would provide more meaningful insights into how query size affects both computational performance and machine learning outcomes.

Additionally, if these larger shifts demonstrated minimal behavioral changes, there would be no need to investigate smaller shifts. However, if the larger shifts produced significant variations in behavior, we could then go back and introduce finer-grained, smaller steps to track patterns more precisely. 

We note that, unlike in a dynamic back-and-forth session with an LLM, where a user may lower their guard or inadvertently reveal PII or pseudo-identifying information, these static queries tend to be less rich in PII. To address this, we manually edited the selected queries to incorporate PII while ensuring the additions remained relevant to the context of the user's original question.

While this manual editing introduces a potential bias, as the PII is not provided by the original author, we believe this approach is more robust than relying on LLM-generated queries. It preserves the integrity of the user-driven nature of the queries while allowing us to test the system's ability to handle sensitive information realistically.
%%%%%%%%%%%%%%%%%%%%%%%%%%%%%%%%%%%%%%%%%%%%%%%%%%%%%%%%%%%%%%%%%%%%%%%%%%%%%%%%%%%%%%%%
\begin{figure}[h!]
  \centering
  % First Subfigure
  \begin{subfigure}[t]{0.35\textwidth} % Adjusted width for side-by-side layout
    \centering
    \includegraphics[width=\textwidth]{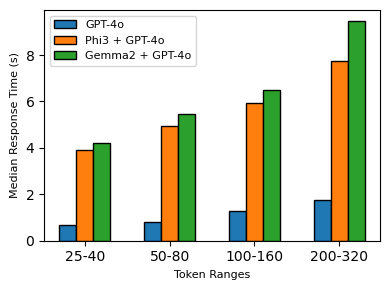}
    \caption{Generic privacy instruction}
    \label{fig:sub1}
  \end{subfigure}
  % \hspace{0.05\textwidth} % Space between the two figures
  % Second Subfigure
  \begin{subfigure}[t]{0.35\textwidth} % Adjusted width for side-by-side layout
    \centering
    \includegraphics[width=\textwidth]{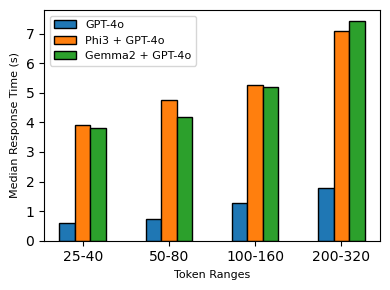}
    \caption{Detailed privacy instruction}
    \label{fig:sub2}
  \end{subfigure}
  
  \caption{Comparison of time taken by the gatekeeper models vs GPT-4.}
  \label{fig:response-time}
\end{figure}

%%%%%%%%%%%%%%%%%%%%%%%%%%%%%%%%%%%%%%%%%%%%%%%%%%%%%%%%%%%%%%%%%%%%%%%%%%%%%%%%%%%%%%%%%%%%%%
\subsubsection{Simulation Performance Measures}
Figure \ref{fig:answer-consistency} illustrates two key measures we track in our simulations. Let $Q$ represent the original user query. When submitted to ChatGPT, it produces an output $A$. The same query is also submitted to our gatekeeper, which refines it into an intermediate query $Q'$, subsequently submitted to ChatGPT, producing a new output $A'$.

We track two important measures of similarity. First, we assess the contextual similarity between $Q$ and $Q'$, which reflects how closely the refined query retains the meaning of the original query. Second, we evaluate the contextual similarity between $A$ and $A'$, measuring how the response generated through our gatekeeper system compares with the direct response from ChatGPT. These measures provide insights into how well our gatekeeper preserves the integrity of both the input query and the final response.

To perform these computations, we use Sentence-BERT (SBERT) \cite{reimers-2019-sentence-bert} to generate embeddings for each pair of texts we aim to compare. We then calculate the cosine similarity between the embeddings to measure their contextual similarity. SBERT is a modification of the pre-trained BERT network that employs a Siamese and triplet network architecture, enabling it to generate semantically meaningful sentence embeddings efficiently. This architecture allows SBERT to maintain BERT's high accuracy while significantly improving the computational efficiency of comparing sentence embeddings \cite{reimers-2019-sentence-bert}.

The final measure we track in our simulations is query response time. Users of our system will want to know how much additional time is incurred due to the dual processing of their queries. If this added time is excessively high, it could disrupt the natural flow of a back-and-forth conversation with the LLM, potentially deterring non-security-conscious users from adopting the technology. To quantify this, we measured the time taken by ChatGPT to return a response and compared it to the time taken by the dual-model approach to process and return a result from the same query.

Another important metric that would be valuable to measure is privacy enhancement: how much our system improves privacy protection. However, due to the lack of an objective, quantifiable metric for privacy suited for our use-case, this analysis is left for the human subjects experiments section where participants will manually assess the responses and provide feedback on any privacy enhancements they notice.  

\subsubsection{Simulation Results}
%%%%%%%%%%%%%%%%%%%%%%%%%%%%%%%%%%%%%%%%%%%%%%%%%%%%%%%%%%%%%%%%%%%%%%%%%%%%%
\label{sec:simulation-results}
Figure \ref{fig:response-time} presents the response times observed during the simulations. We chose to use the median response times rather than the mean, as occasional outliers, likely caused by network issues when queries were sent via the ChatGPT API, would have skewed the mean. Since these outliers appeared randomly and were not a characteristic of the system under investigation, we opted to mute their impact.

The figure shows a consistent increase in response time as token lengths grow, for both the dual-model architecture and GPT-4o. It also demonstrates that GPT-4o consistently delivers faster responses. These trends hold true for both generic and detailed privacy instructions. It's important to note, however, that the increased delays remain within the range of a few seconds. User experiments with the prototype will provide insight into how participants perceive these delays, especially given the potential trade-off for enhanced privacy.

One notable observation is that the smaller model (Phi3) consistently performs slightly better than the larger model (Gemma2) when using generic instructions. However, with detailed instructions, this trend reverses. That said, the differences between the two models are small, and likely not statistically significant. We plan to investigate this phenomenon further in future work.

\begin{figure}[h!]
  \centering
  % First Subfigure
  \begin{subfigure}[b]{0.35\textwidth} % Adjusted to fit the column
    \centering
    \includegraphics[width=\columnwidth]{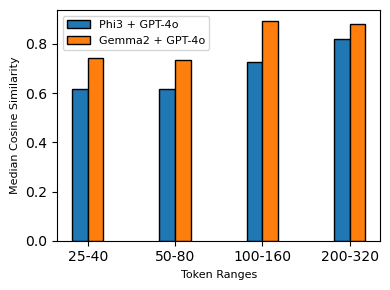}
    \caption{Generic privacy instruction}
    \label{fig:sub1}
  \end{subfigure}
  % Second Subfigure
  \begin{subfigure}[b]{0.35\textwidth} % Adjusted to fit the column
    \centering
    \includegraphics[width=\columnwidth]{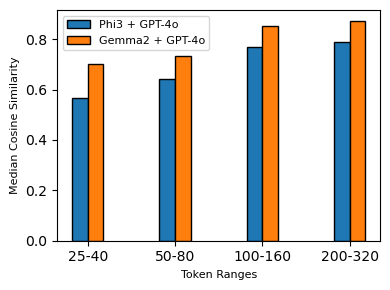}
    \caption{Detailed privacy instruction}
    \label{fig:sub2}
  \end{subfigure}
  \caption{Cosine similarity between input and refined queries.}
  \label{fig:cosine-sim-input-refined}
\end{figure}

Figure \ref{fig:cosine-sim-input-refined} shows the semantic similarity between the original query and the refined query. As before, we used median values to mitigate the effect of random cases of model hallucination. The figure demonstrates that the larger model consistently outperforms the smaller model across all token sizes and query lengths. However, both models retain substantial semantic meaning in the refined queries, with the similarity increasing as token length grows. Note that a cosine similarity of 1 indicates perfect similarity, while -1 represents opposite meanings.
    
%%%%%%%%%%%%%%%%%%%%%%%%%%%%%%%%%%%%%%%%%%%%%%%%%%%%%%%%%%%%%%%%%%%%%%%%%%%%%%%%%%%%%%%%%%%%%%

\begin{figure}
  \centering
  % First Subfigure
  \begin{subfigure}[b]{0.35\textwidth}
    \centering
    \includegraphics[width=\columnwidth]{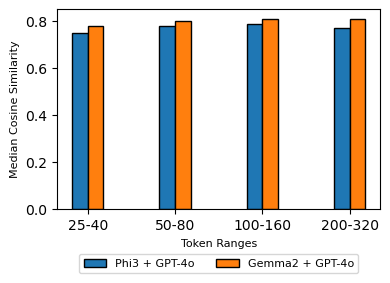}
    \caption{Generic privacy instruction}
    \label{fig:cosine-mixed-generic}
  \end{subfigure}
  % Second Subfigure
  \begin{subfigure}[b]{0.35\textwidth}
    \centering
    \includegraphics[width=\columnwidth]{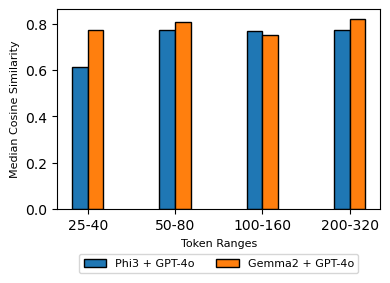}
    \caption{Detailed privacy instruction}
    \label{fig:cosine-mixed-detailed}
  \end{subfigure}  
  \caption{Cosine similarity between mixed model output and GPT-4o output}
  \label{fig:cosine-sim-mixed-model}
\end{figure}

  Figure \ref{fig:cosine-sim-mixed-model} presents the semantic similarity evaluation between the responses generated by our dual-model architecture and GPT-4. As observed earlier, the larger model consistently outperforms the smaller one, though the difference here is more marginal compared to the previous figure. Interestingly, the smaller model seems to recover somewhat in the final output, suggesting that the lower cosine similarity scores seen in the intermediate step do not necessarily translate into comparatively poorer performance in the final result. This observation is particularly noteworthy and warrants further investigation into the interaction between the intermediate refinement and the final response quality.
  
  %%%%%%%%%%%%%%%%%%%%%%%%%%%%%%%%%%%%%%%%%%%%%%%%%%%%%%%%%%%%%%%%%%%%%%%%%%%%%%%%%%%%%%%%%%%%%%

\subsection{Experiment with Human Subjects}
\label{sec:experiments-human-subjects}
Following approval by our university's Institutional Review Board (IRB), we recruited human subjects to participate in our experiment. Below, we provide the details of the experiment.
\subsubsection{User Interactions with our Prototype}
\label{sec:user-interaction-prototype}
Our experiment involved 39 participants interacting with our prototype in a controlled lab environment. All participants were students and staff from our university and had a basic working knowledge of language models such as ChatGPT. Each participant submitted three health-related queries per gatekeeper model, resulting in six assessments per participant.

Participants were instructed to include fictitious private information in their queries, fully aware that they would be evaluating how well this information was filtered out. The experiment flow was as follows: each participant entered a query, selected a gatekeeper model, received a response, and then completed a survey about that interaction before moving on to the next query.

The assessment questions are outlined in Table \ref{tab:survey-2}, focusing on four key areas: the effectiveness of privacy enhancement, the relationship between the refined query and the original query, how well the final answer met the participant's expectations, and their impressions of the response time.

%After completing their interaction, they provided feedback on the prototype. 

% Please add the following required packages to your document preamble:
% \usepackage{graphicx}
% Please add the following required packages to your document preamble:
% \usepackage{graphicx}
\begin{table}[]
\centering
\resizebox{\columnwidth}{!}{%
\begin{tabular}{rl}
\hline
\textbf{Q\#}  & \multicolumn{1}{c}{\textbf{Questions}}                                                                                                                                                                                          \\ \hline
\textbf{Q8}   & \begin{tabular}[c]{@{}l@{}}I am convinced that the refined query generated by the intermediate\\ model filtered out my private information.\end{tabular}                                                                         \\ \hline
\textbf{Q9}   & \begin{tabular}[c]{@{}l@{}}I am convinced that the refined query generated by the intermediate\\ model preserved the meaning of my original query.\end{tabular}                                                                  \\ \hline
\textbf{Q10}   & \begin{tabular}[c]{@{}l@{}}Based on the final response generated by the model, I feel that\\ the model understood my question.\end{tabular}                                                                                      \\ \hline
\textbf{Q11}   & \begin{tabular}[c]{@{}l@{}}I feel that the delay between posting my question and receiving\\ the response was acceptable (i.e., response time was not abnormally long).\end{tabular}                                              \\ \hline
\end{tabular}%
}
\caption{Survey questions answered by users after interacting with our system}
\label{tab:survey-2}
\end{table}

% \begin{figure*}[htbp]
%   \centering
%   % First Subfigure
%   \begin{subfigure}[t]{0.3\textwidth}
%     \centering
%     \includegraphics[width=\textwidth]{Q1&Q4.png}
%     \caption{Responses for Q1 and Q4}
%     \label{fig:q1-q4}
%   \end{subfigure}
%   \hspace{0.02\textwidth} % Space between figures
%   % Second Subfigure
%   \begin{subfigure}[t]{0.3\textwidth}
%     \centering
%     \includegraphics[width=\textwidth]{Q2&Q3.png}
%     \caption{Responses for Q2 and Q3}
%     \label{fig:q2-q3}
%   \end{subfigure}
%   \hspace{0.02\textwidth} % Space between figures
%   % Third Subfigure
%   \begin{subfigure}[t]{0.3\textwidth}
%     \centering
%     \includegraphics[width=\textwidth]{Q5.png}
%     \caption{Responses for Q5}
%     \label{fig:q5}
%   \end{subfigure}

%   \caption{Responses across survey questions 1 to 5.}
%   \label{fig:main}
% \end{figure*}

\subsection{Results from Human Subjects Experiment}
% \subsubsection{User Survey Result}
\label{sec:user-study-results}

\subsubsection{Results from User Interactions with our Prototype}

\begin{figure}[h!]  % Using figure* to span both columns
  \centering
  % First Subfigure
  \begin{subfigure}[t]{0.35\textwidth}
    \centering
    \includegraphics[width=\textwidth]{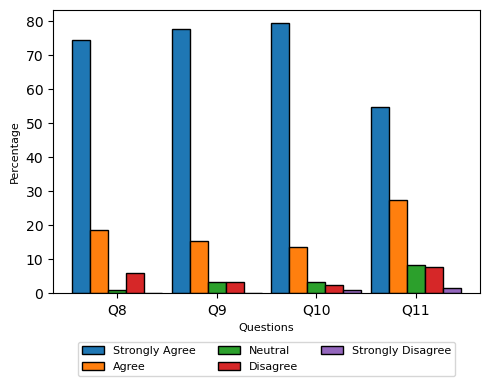}
    \caption{Phi3 + GPT-4o}
    \label{fig:user-assessment-phi3}
  \end{subfigure}
  \hspace{0.05\textwidth} % Space between figures
  % Second Subfigure
  \begin{subfigure}[t]{0.35\textwidth}
    \centering
    \includegraphics[width=\textwidth]{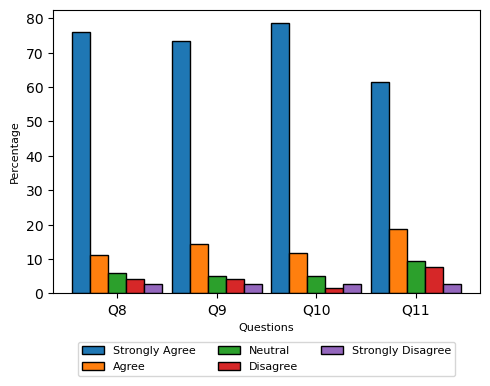}
    \caption{Gemma2 + GPT-4o}
    \label{fig:user-assessment-gemma2}
  \end{subfigure}
  
  \caption{User responses following interactions with Phi3 + GPT-4o and Gemma2 + GPT-4o models.}
  \label{fig:user-assessment}
\end{figure}

Figure \ref{fig:user-assessment} presents the results from the survey conducted after users interacted with our system. The survey questions, numbered 8 to 11, are detailed in Table \ref{tab:survey-2}. The figure reveals a clear pattern: users strongly agreed that the system effectively filtered out their private data, the refined query retained its original meaning, the model accurately understood their questions, and the response delay was acceptable. This positive feedback was consistent across both gatekeeper models. Of the four aspects, response time consistently received slightly weaker feedback, though it was still rated acceptable by most participants. The results suggest that while users generally appreciated the privacy-enhancing features and the accuracy of the system, response time remains a potential area for improvement. 

%%%%%%%%%%%%%%%%%%%%%%%%%%%%%%%%%%%%%%%%%%%%%%%
% survey part1 response

% % First Figure
% \begin{figure*}[t]
%   \centering
%   \includegraphics[0.20\columnwidth]{sensitivedata.png}
%   \caption{Q6: Sensitive Data Types and Concerned Percentages}
%   \label{fig:q6}
% \end{figure*}
% \hfill

% \subsection{User Feedback and Perceptions}

% \label{sec:user-feedback}
% \begingroup

% % First Set of Figures
% \begin{figure*}[h]  % Using figure* to span both columns
%   \centering
%   % First Figure
%   \begin{minipage}[t]{0.45\textwidth}
%     \centering
%     \includegraphics[width=\textwidth]{sensitivedata.png}
%     \caption{(Q6) Sensitive Data Types}
%     \label{fig:q6}
%   \end{minipage}
%   \hspace{0.05\textwidth} % Space between figures
%   % Second Figure
%   \begin{minipage}[t]{0.45\textwidth}
%     \centering
%     \includegraphics[width=\textwidth]{misuse.png}
%     \caption{(Q7) Perceptions of Potential Data Misuse}
%     \label{fig:q7}
%   \end{minipage}
% \end{figure*}
%%cutting out the above for ICSC%%%%
%%%%%%%%%%%%%%%%%%%%%%%%%%%%%%%%%%%%%%%%%%

% Second Set of Figures
% \FloatBarrier % Prevents figures from floating into the references section
%%%%%%%%%%%%%%%%%%%%%%%%%%%%%%%%%%%%%%%%%%%%%%%%%%%%%%%%%%%%%%
\section{Discussion and Conclusions}
\label{sec:discussion-conclusions}
In this paper, we conducted a user survey on various LLM privacy issues and introduced an LLM gatekeeper system aimed at addressing the privacy risks posed by sharing sensitive information with cloud-based language models.
% The survey results indicate that users generally distrust LLM providers when it comes to protecting their data and are concerned that their information could be misused in various ways, even when AI providers claim not to misuse it. 
This highlights the pressing need for robust privacy measures like the system we propose, which not only addresses privacy concerns but also reassures users, giving them confidence that their sensitive information is safeguarded before it is sent to third-party platforms.

Our system demonstrates that real-time filtering of PII is achievable without significantly compromising the quality of user interactions. Through both simulations and a user study, we showed that our approach not only enhances privacy but also preserves the semantic meaning of user queries, making it practical for everyday use.

Despite these successes, there are areas for improvement. One of the main challenges is minimizing the response time introduced by the dual-model architecture, especially for users in scenarios requiring quick, conversational back-and-forth exchanges. Reducing this response time is critical to maintaining a natural and seamless interaction flow.

Future work could explore the use of smaller, specialized models optimized solely for PII filtering to reduce computational overhead. Additionally, advancements in hardware, such as GPUs and dedicated AI accelerators, could further improve the system’s efficiency. By leveraging these techniques, it might be possible to strike a better balance between privacy and performance, enabling the adoption of privacy-enhancing systems that not only protect users' data but also meet the growing demands for responsive and efficient AI interactions.

% The next two lines define the bibliography style to be used, and
% the bibliography file.
\bibliographystyle{IEEEtran}
\bibliography{main} %, mybibfile}
\end{document}